\documentclass[journal]{IEEEtran}

\usepackage{graphicx}
\usepackage{amsmath}
\usepackage{amsfonts}
\usepackage{cite}
\usepackage{subfigure}

\begin{document}

\title{Optimum Tag Reading Efficiency of Multi-Packet Reception Capable RFID Readers}

\author{Subodh Pudasaini, Seokjoo Shin, and Kyung Sup Kwak % <-this % stops a space
\thanks{S. Pudasaini and K. S. Kwak are with UWB-Wireless Communications Research Center, Inha University, Incheon, Korea.}
\thanks{S. Shin is with Department of Computer Engineering, Chosun University, Gwangju, Korea}
}
%\markboth{IEEE Communications Letters}%
%{Author \MakeLowercase{\textit{et al.}}: Optimum Tag Reading Efficiency of Multi-Packet Reception Capable RFID Readers}



\maketitle

\begin{abstract}
Maximizing the tag reading rate of a reader is one of the most important design objectives in RFID systems as the tag reading rate is inversely proportional to the time required to completely read all the tags within the reader’s radio field. To this end, numerous techniques have been independently suggested so far and they can be broadly categorized into pure advancements in the link-layer tag anticollision protocols and pure advancements in the physical-layer RF transmission/reception model. In this paper, we show by rigorous mathematical analysis and Monte Carlo simulations that how such two independent approaches can be coupled to attain the optimum tag reading efficiency in a RFID system considering a Dynamic Frame Slotted Aloha  based link layer anti-collision protocol at tags and a Multi-Packet Reception capable RF reception model  at the reader.
\end{abstract}


% \fi
\IEEEpeerreviewmaketitle

\section{Introduction}
\IEEEPARstart{R}{adio} Frequency Identification (RFID) is a rapidly evolving automatic identification and tracking system. Even though the basic operating principles of modern RFID systems have been known for several decades, their adoption in numerous industrial and consumer applications (such as supply chain management, inventory control, supermarket checkout process, and toll collections)  has been proliferated recently due to the ability now to build miniaturized RFID components at low cost \cite{landt}.

Typically, a RFID system consists of two components: a reader and tags. Each tag has a unique ID stored in its memory. The reader should read (interrogate) IDs of all the tags within its radio field, and for this purpose it broadcasts interrogation RF signal periodically. If an RFID tag finds itself within the RF-field of the reader, it backscatters (i.e. transmits back) a signal containing its unique ID \cite{rao}. When more than one RFID tags backscatter their IDs using a common chunk of the shared wireless channel (in terms of frequency, time, space, or code), signal from one tag interferes the signals from others, and the reader might not be able to decode IDs of the backscattering tags. Such phenomenon is commonly known as tag-collision. Occurrence of such tag-collision events triggers the collided tags to retransmit their IDs in the subsequent interrogation rounds and thus elongates tag identification delay (or in other words reduces the tag reading rate) at the reader. Many link-layer (more precisely medium access control sub-layer) anti-collision protocols have been developed so far to address the tag-collision problem \cite{klair}. Those protocols not only reduce the frequency of occurrence of tag-collision events but also help to recover from such events as quickly as possible.

In a broad sense, time division multiple access RFID anti-collision protocols are classified as either deterministic or probabilistic protocols based on how tags are allocated a fraction of the shared channel resource (a time slot) to transmit their IDs. The former type of protocols is based on Binary Tree (BT) where the collided tags are split into two subsets. The tags in the first subset transmit their IDs in the next slot, while the tags in the other subset wait until the first subset of tags are successfully identified. This process is repeated recursively until all tags are recognized. The performance of tree-based anticollision protocols deteriorates with increase in the number of tags. This is because even though the colliding tags are successively grouped into two subsets, each subset may still contain many tags resulting in collision \cite{myung}. On the other hand, in probabilistic protocols such as Framed Slotted Aloha (FSA), the channel time is split into frames and a single frame is further divided into several time slots. During each frame, each tag randomly chooses a time slot and transmits its ID to the reader in that slot. The unidentified tags will transmit their IDs in the next frame. It has been shown that the probabilistic FSA can achieve smaller tag identification delay than its deterministic counterpart \cite{chen1}.

In the literature there exist many works which have been independently developed by different researchers and engineers to enhance the tag identification performance of a RFID system. Some of the representative works are available in \cite{vogt,khandelwal,mindikoglu,dacuna}. Based on the scope of their design, they can be categorized into (i) pure advancement in the link-layer anticollision protocols, and (ii) pure advancement in the physical layer RF transmission/reception models. The fundamental approach behind the first category of enhancements is to dynamically adjust the frame length of the probabilistic FSA protocols to its optimal value in each interrogation round (resulting in new protocol referred to as Dynamic FSA or DFSA \cite{vogt}), or to optimize tree search algorithm in the deterministic BT protocols taking advantage of inherent correlatedness among the tag IDs \cite{khandelwal}. The latter category of enhancement, on the other hand, uses Multiuser-Multiple Input Multiple Output (MU-MIMO) technique along with efficient blind signal separation algorithms to realize Multi-Packet Reception (MPR) capable RF reception model at the reader \cite{mindikoglu,dacuna}. Due to the MPR capability at reader, simultaneously transmitted signals from several tags can be separated and the transmitting tags can be correctly identified (which otherwise would have been treated as being collided).

It has been shown in \cite{lee, kim} that MPR capability at reader has potential to substantially increase the read rate and decrease identification delay of FSA and BT anticollision protocols, respectively. However, how to ascertain optimal tag reading performance in a RFID system with MPR capability is remained as an open research problem. To this end, we derive an optimality criterion and present a method to adopt such a criterion in the probabilistic DFSA anticollision protocol in a RFID system with MPR capability. To the best of our knowledge, it is the first work in this regard.

The rest of the paper is organized as follows. Section II presents the system model while Section III presents analytical derivation of a criterion for achieving optimal tag reading efficiency. Section IV provides detail information about simulations environment, performance metrics and evaluation methodology. Finally, Section V concludes this work.

\section{System Model}
 We consider a RFID system where $n$ number of tags with single antenna communicates with a reader equipped with array of $M$ antennas. Under such MU-MIMO setting, it is assumed that spatially multiplexed backscattered signals from multiple tags can be separated at the reader using advanced signal processing algorithms unless the number of multiplexed signals exceeds $M$.  Dacuna et al \cite{dacuna} have recently demonstrated the feasibility of such assumption in UHF RFID systems.

DFSA is used as the anticollision protocol. The operation procedures of DFSA at the reader and tag are described below:\\
\noindent \textbf{Reader side:} (1) Set initial frame length. (2) Initiate interrogation round by broadcasting the frame length information. (3) In each slot of the frame, check whether there are any backscattered RF signals from the tags. Mark the slot as an empty slot if no backscattered RF signal is detected. If RF signals are detected, use the advanced signal separation algorithm to separate the multiplexed backscatter RF signals. Based on the outcome of the signal separation  operation, mark the slot as a  collided slot if none of the transmitting tags are identified, and mark it as a successful slot if any of the tags are identified. Also record the number of identified tags in the successful slot. (4) After the completion of the frame, check whether any slot within that frame is marked as the collided slot. It is the indication whether any tags are left to be interrogated or not. If none of the slots are marked as the collided slot, terminates the interrogation process. Otherwise, prepare for the next interrogation round. (5) Estimate the total number of contending tags in the last frame using \emph{maximum a posteriori} based estimation method in Eq. (11). As to be elaborated in the next section, the MAP estimation mechanism utilizes the statistics of the collided, successful, and idle slots to perform estimation. (6) Determine the optimal frame length for next interrogation round using Eq. (12) and go to step (2).\\

\noindent \textbf{Tag side:} (1) Wait for interrogation signal from the reader. (2) Obtain the frame length information. (3) Randomly select any of the slot within the frame and backscatter its ID in the selected slot. (4) If the transmission is inferred to be unsuccessful, wait for interrogation signal for the next round.

\section{Optimal Tag Reading Criterion}

In this section, we derive a theoretical criterion for achieving optimal tag reading performance at the reader with MPR capability and present a method to use such criterion in the practical RFID systems.

Consider the RFID system described in the previous section with $n$ tags to be read.  The frame used in an interrogation round initiated by the reader consists of $L$ time slots. So, the probability that $j$ tags among $n$ tags occupy a slot can be expressed by the binomial distribution with parameters $n$ and $1/L$ as
\begin{equation}
B(j)=\binom{n}{j}\left(\frac{1}{L}\right)^j\left(1-\frac{1}{L}\right)^{n-j}.
\end{equation}
\noindent If the frame length $L$ is sufficiently large, Eq. (1) can be approximated by the Poisson distribution with mean $n/L$. Accordingly, the probabilities that a slot is found to be empty (no tags use the slot), successful ($M$ or less number of tags use the slot), and collided (more than $M$ number of tags use the slot) are given by
\begin{eqnarray}
p_e&=&B(j=0)\approx e^{-n/L}, \\
p_s&=&B(1\leq j \leq M) \approx e^{-n/L}\sum_{j=1}^{M}\frac{(n/L)^j}{j!}, \ \ \text{and} \\
p_c&=&B(j>M)=1-p_e-p_s.
\end{eqnarray}
\noindent Based on Eq. (3), the expected value of the number of successful slots in the frame with $L$ slots is
\begin{equation}
E[S]=L\cdot e^{-n/L}\sum_{j=1}^{M}\frac{(n/L)^j}{j!}.
\end{equation}
\noindent To maximize read rate (number of successful tags per unit time) of the reader it should be ensured that the shared channel should be used as efficiently as possible. This implies that a criterion that maximizes the channel usage efficiency $U$ (defined as a ratio of expected value of the number of successful slots to the frame length) also maximizes the read rate. Since $U$ is a concave-downward function of $L$, the criterion that maximizes $U$ can be obtained by equating the derivative of $U$ with respect to $L$ to zero as
\begin{eqnarray}
\frac{dU}{dL}&=&\frac{d}{dL}\left(e^{-n/L}\frac{n}{L}\right)+\frac{d}{dL}\left(e^{-n/L}{(\frac{n}{L})^2}\frac{1}{2!}\right)+ \cdot \cdot \cdot \\ \nonumber
 &+&\frac{d}{dL}\left(e^{-n/L}{(\frac{n}{L})^M}\frac{1}{M!}\right)=0.
\end{eqnarray}
\noindent Further simplification of Eq. (6) yields
%\begin{eqnarray}
%\frac{n}{1!L^2}e^{-n/L}\left(\frac{n}{L}-1\right)+\frac{n^2}{2!L^3}e^{-n/L}\left(\frac{n}{L}-2\right)+\cdot \cdot %\cdot \\ \nonumber  + \frac{n^M}{M!L^{M+1}}e^{-n/L}\left(\frac{n}{L}-M\right)= 0
%\end{eqnarray}

\begin{equation}
\sum_{m=1}^{M}\left(\frac{n^m}{m!L^{m+1}}e^{-n/L}\left(\frac{n}{L}-m\right)\right)=0.
\end{equation}
\noindent Solving Eq. (7), the criterion (i.e. optimal frame length $L^*$) that maximizes $U$ is found to be
\begin{equation}
L^*=\frac{n}{(M!)^{\frac{1}{M}}}.
\end{equation}
\noindent If the number of tags to be interrogated is known in advance, a value of the frame length for the optimal usage of   DFSA can be set to the value obtained from Eq. (8). However, the cardinality of tags to be interrogated is not known in advance. Hence, for each frame, except for the initial frame, remaining tags to be interrogated should be estimated on-the-fly.

Chen has previously proposed a Maximum a Posteriori (MAP)-based tag estimation method \cite{chen2} and showed that it is more accurate than its predecessors such as Vogt’s method \cite{vogt}. Chen however did not consider MPR capabilities in the reader and hence his tag estimation formula is applicable for single packet reception model only (i.e. $M = 1$). In what follows, we extend Chen’s formula for all possible values of $M$. In a frame with $L$ slots, the joint probability mass function for finding $X$ empty slots, $Y$ successful slots and $Z$ collision slots can be represented using the following trinomial distribution
\begin{equation}
P(X,Y,Z)=\frac{L!}{X!Y!Z!}p_e^Xp_s^Yp_c^Z,
\end{equation}
\noindent where $p_e$, $p_s$ and $p_c$ are previously defined in Eq. (2), (3), and (4), respectively. Hence, when the reader finds $E$ empty slots, $S$ successful slots, and $C$ collision slots in a frame, a posteriori probability distribution of having $k$ tags in the system is
\begin{eqnarray}
P(k|E,S,C)&=&\frac{L!}{E!S!C!} \left(e^{-k/L}\right)^E \notag\\
&\times & \left[e^{-k/L}\left(T_M(k/L)-1\right)\right]^S \notag \\
&\times & \left[e^{-k/L}\left(e^{k/L}-T_M(k/L)\right)\right]^C,
\end{eqnarray}
\noindent where $T_M(k/L)$ is the Taylor polynomial of $e^{k/L}$ of order $M$. Based on the \emph{posterior }probability distribution in Eq. (10), the reader determine the total number of estimated tags as
\begin{equation}
\hat n = \arg\!\max\limits_{k} P(k|E,S,C).
\end{equation}
\noindent Once the number of tags in $(i-1)$th frame is estimated using Eq. (11), the optimal frame length in the next frame for interrogating the remaining tags will be
\begin{equation}
L^*_i=\frac{\hat n_{i-1}-S_{i-1}}{(M!)^{\frac{1}{M}}}, \ i> 1,
\end{equation}
\noindent where $S_{i-1}$ is the number of successfully identified tags in the $(i-1)$th frame.

Fig. 1 shows the posteriori probability distribution for $n$ tags when 1 empty slot, 6 collision slots, and 3 success slots are observed in a frame with 10 slots for three different cases of $M$ (viz. $M = 1$, $M = 2$ and $M = 3$). For each case of $M$, the value corresponding to the peak of the distribution curve is the estimated number of tags.

 \begin{figure}[h]
\centering
\includegraphics[scale = .75]{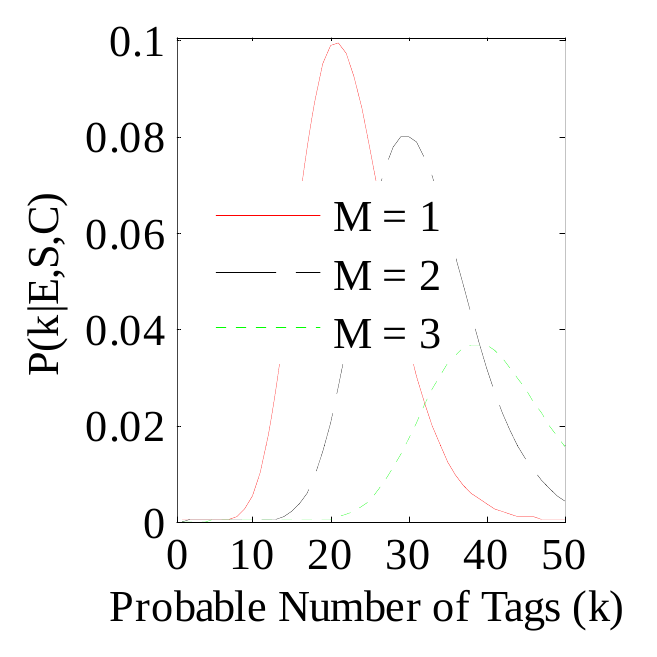}
\caption{\small \emph{Posteriori} probability distribution of estimated number of tags for different values of $M$ when $ L = 10$, $C = 6$, $S = 3$, and $E = 1$.}
\end{figure}

It is noteworthy to mention that while implementing the MAP-based estimation method in the reader, the first constant factor (involving factorial) in $P(k│E,S,C)$ can be removed as it is only responsible in scaling the probability mass function. There will be no difference in the estimation result but significant computation burden from the reader can be reduced, especially when $L$ is large.

\section{Performance Results}
We analyzed the performance of the MPR capable RFID system described in Section II for varying $M$, $L$ and $n$ using Monte Carlo simulations. Average results of 500 simulation trials are presented in terms of two metrics defined below: (a) Read rate: Number of tags identified per unit time, and (b) Identification delay: Total time required to read all the tags in the system. We considered the duration of a slot to be a basic unit of time, and hence the read rate is expressed in terms of tag/slot (number of tags per slot) and identification delay in terms of number of slots.

Fig. 2a (left) shows read rate of a FSA anticollision algorithm and DFSA anticollision algorithm with varying MPR capabilities ($M =$ 1, 2, 3 and 4) when the initial frame length was set to 128.  It is evident from the figure that read rate substantially increases with increase in the value of $M$. This is attributed to the reduction in the number of tag-collision events due to MPR capability. Read rate reaches its peak value of 1.9 tags/slot for the case of $M =$ 4, which in the conventional single packet reception capable reader (i.e., $M =$ 1) is caped to 0.36 tag/slot. Note that DFSA’s peak read rate in the single packet reception capable reader agrees well to the previously established theoretical network throughput bound of $\frac{1}{e}$ ($\approx$ 0.37) in any aloha based random access systems. In the figure, it is also evident that by merely using FSA it is not possible to attain the read rate closer to $\frac{1}{e}$ in the single packet reception capable reader.

Fig. 2a (right) shows the identification delay of FSA anticollision algorithm and DFSA anticollision algorithm with varying MPR capabilities. From the figure one can see that the increased read rate due to MPR capabilities (observed in Fig. 2a (right)) translates to the reduction in the identification delay. For example, when there were around 350 tags in the RFID system, nearly 5.5 fold decrease in the identification delay (from 1011 slots to 184 slots) was observed when the single packet reception capable reader was replaced with MPR-capable reader with $M =$ 4.

Fig. 2b shows that the initial frame length affects  the performance of DFSA both in terms of read rate and identification delay, especially when the reader has high-order MPR capabilities and the number of tags to be interrogated is small. From the figure it is evident that the read rates for three different cases of initial frame lengths ($L =$ 32, 64 and 128) appear to converge to a rate close to the peak read rate with increase in the number of tags in the system. This implies that the effects of the initial frame length on read rate tends to vanish with increase in the number of tags. Similarly, the difference in identification delay for different frame length values shrinks for larger number of contending population size.

Next, we measured the accuracy of the MAP-based tag estimation method used in our previous simulations. For that we calculated the estimation error (in \%) as $\frac{|\hat n-n|}{n}\times 100\%$, where where $\hat n$ is the estimated number of tags when there were $n$ tags in the system.  The lower value of the estimation error corresponds to the higher estimation accuracy. Fig. 3 depicts estimation errors for four different cases of $M$ (1, 2, 3, and 4) when the frame length was set to 128 in the simulations. From the figure it is evident that the estimation error increases with increase in the value of $M$, but only up to a certain tag population size. Beyond that tag population size, estimation error for higher $M$ remains lower. Importantly, for all four different cases of $M$, the estimation error remains lower than 6\% regardless of the number of considered tags.

\begin{figure}[t]
\centering
\vspace{2mm}
\subfigure[]{
\includegraphics[scale = .75]{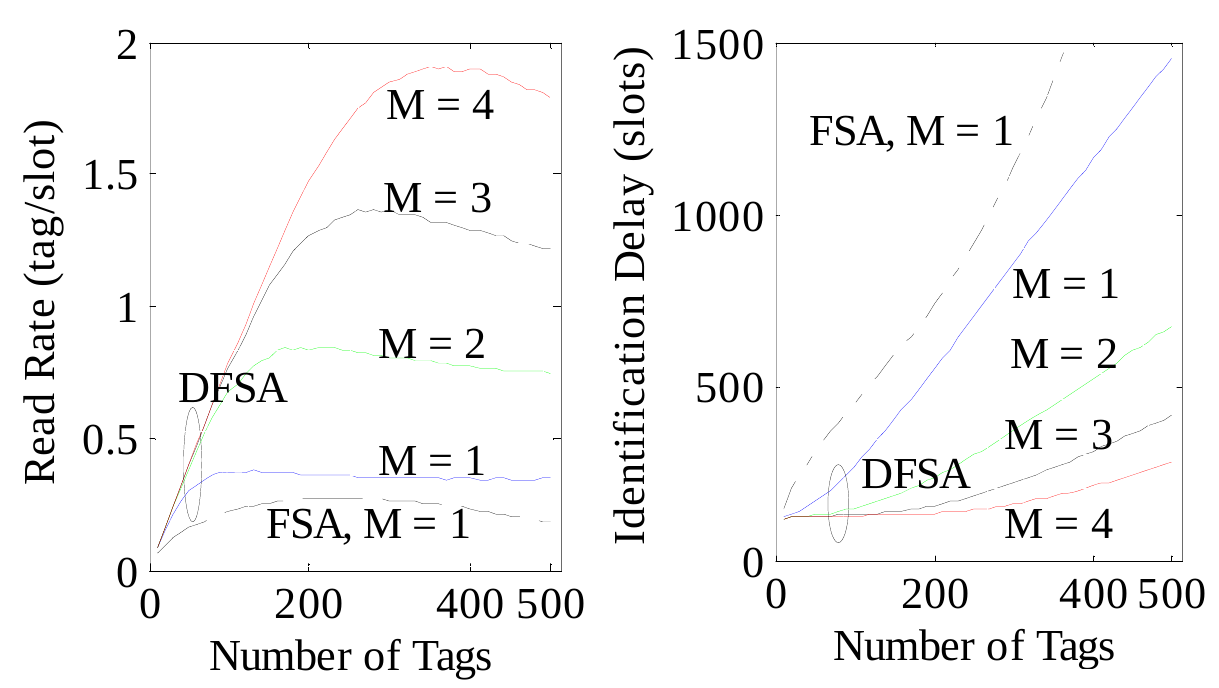}
\label{fig:2a}}
\subfigure[]{
\includegraphics[scale = .75]{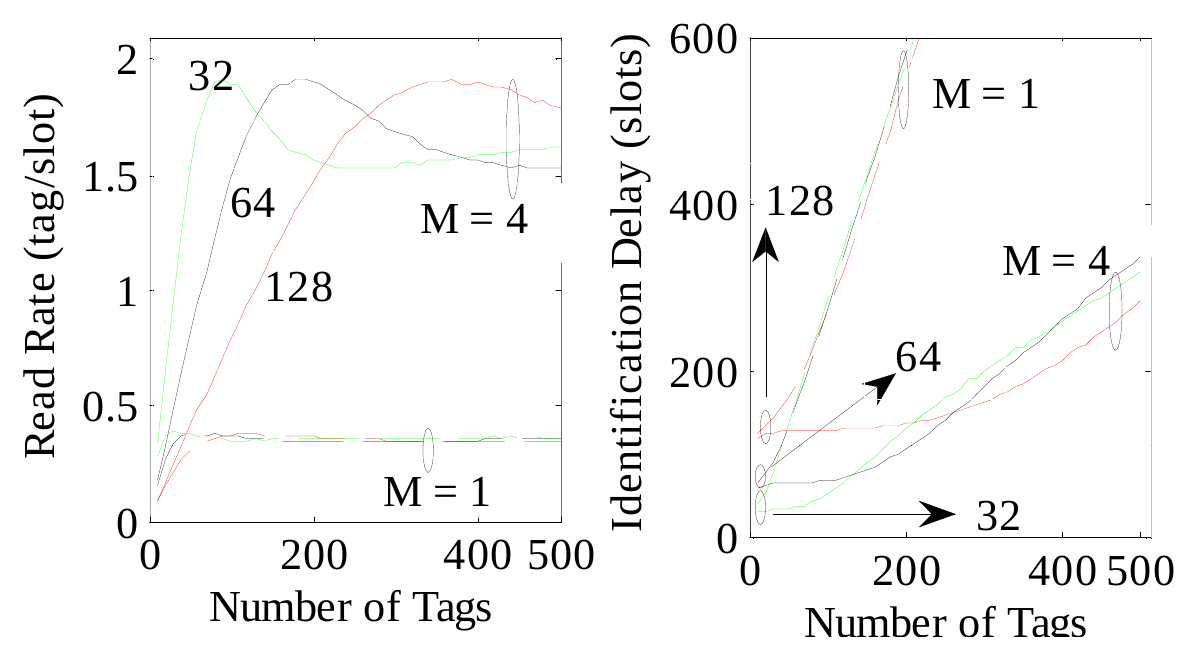}
\label{fig:2b}}
\caption{\small Read rate and identification delay of DFSA anticollision protocol in a RFID system with multi packet reception capable reader (a) Influence of $M$, (b) Influence of initial $L$}
\end{figure}

 \begin{figure}[h]
\centering
\vspace{2mm}
\includegraphics[scale = .75]{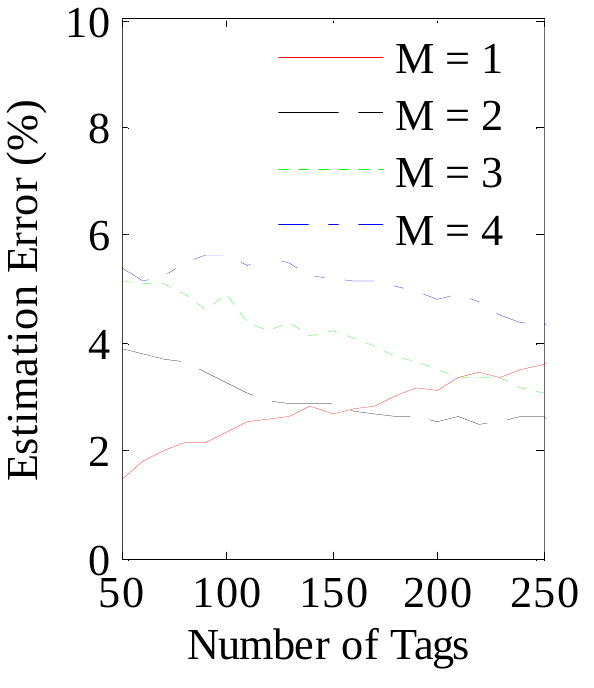}
\caption{\small Difference between the real number of tags and the estimated number of tags (expressed in percentage) when $L$ was set to 128.}
\end{figure}

\section{Conclusion}
In this paper, we have derived a general criterion to achieve the optimal performance of a probabilistic DFSA based anticollision algorithm in RFID system with MPR capable reader. Previously, only the criterion for the single packet reception capable reader was known. Further, we have provided a simple method to adopt such a criterion in practical  RFID systems. Through rigorous computer simulations, we have shown the performance implications of that optimal criterion in terms of increased tag reading rate and reduced identification delay.

\ifCLASSOPTIONcaptionsoff
  \newpage
\fi



%\bibliographystyle{IEEEtran}

\end{document}